\documentclass[reprint, %
superscriptaddress,
amsmath,amssymb, %
aps, %
]{revtex4-1}
\pdfoutput=1
\usepackage{hhline}
\usepackage{xcolor}
\usepackage{hyperref}
\usepackage{graphicx}
\usepackage{bm}
\numberwithin{equation}{section}

\begin{document}

\title{Coupled quintessence with a $\Lambda$CDM background: removing the $\sigma_8$ tension}

\author{Bruno J. Barros}
\affiliation{Instituto de Astrof\'isica e Ci\^encias do Espa\c{c}o,\\ 
Faculdade de Ci\^encias da Universidade de Lisboa,  \\ Campo Grande, PT1749-016 
Lisboa, Portugal}
\author{Luca Amendola}
\affiliation{Institut f\"ur Theoretische Physic, Universit\"at Heidelberg, Philosophenweg 
16, D-69120 Heidelberg, Germany}
\author{Tiago Barreiro}
\affiliation{Departamento de Matem\'atica,  ECEO, Universidade Lus\'ofona de Humanidades e 
Tecnologias, Campo Grande, 376,  1749-024 Lisboa, Portugal}
\affiliation{Instituto de Astrof\'isica e Ci\^encias do Espa\c{c}o,\\ 
Faculdade de Ci\^encias da Universidade de Lisboa,  \\ Campo Grande, PT1749-016 
Lisboa, Portugal}
\author{Nelson J. Nunes}
\affiliation{Instituto de Astrof\'isica e Ci\^encias do Espa\c{c}o,\\ 
Faculdade de Ci\^encias da Universidade de Lisboa,  \\ Campo Grande, PT1749-016 
Lisboa, Portugal}

\date{\today}

\begin{abstract}
A well-known problem of the $\Lambda$CDM model is the tension between the relatively high level of clustering, as quantified by the parameter 
$\sigma_8$, found in cosmic microwave background experiments and the smaller one obtained from large-scale observations in the late Universe. In this paper we show that coupled quintessence, i.e. a single dark energy scalar field conformally coupled to dark matter through a constant coupling,
can solve this problem if the background is taken to be identical to the 
$\Lambda$CDM one. We show that two competing effects arise. On one hand,  the additional scalar force is attractive, and is therefore expected to increase the clustering. On the other,  in order to obtain the same background as 
$\Lambda$CDM, coupled quintessence  must have a smaller amount of dark matter near the present epoch. We show that the second effect is dominating today and leads to an overall slower growth. Comparing to redshift distortion data, we find that coupled quintessence with $\Lambda$CDM background solves the tension between early and late clustering.

We find for the coupling $\beta$ and for $\sigma_8$ the best fit values
$|\beta| = 0.079^{+ 0.059}_{- 0.067}$ and
$\sigma_8 = 0.818^{+0.115}_{-0.088}$. These values also fit the lensing data from the KiDS-450 survey. We also estimate that the future missions SKA and Euclid will constrain 
 $\beta$ with an error of $\pm\, 1.5\times10^{-3}$ and for $\sigma_8$ of
  $\pm \,1.8\times10^{-3}$ at $1\sigma$ level.
\end{abstract}

\maketitle

\section{Introduction}
\label{introduction}

The recent realization that  the Universe expansion is accelerating \cite{Perlmutter:1998np,Riess:1998cb} has puzzled cosmologists to this day and has lead them to conjecture the existence of a so called dark energy \cite{Copeland:2006wr}. The standard model of cosmology, the $\Lambda$CDM model, describes dark energy as a cosmological constant $\Lambda$ \cite{Peebles:2002gy}
 with a constant energy density and negative pressure, fighting the gravitational pull of matter and causing the expansion to accelerate. 
Even though the $\Lambda$CDM model presents a good fit to the present observations (particularly at the background level), it presents some conceptual problems \cite{Peebles:2002gy,Weinberg:2000yb} motivating us to explore other possibilities for the dark sector. One enticing possibility is a form of dynamical dark energy \cite{Wetterich:1994bg} in which the acceleration is induced by a scalar field, usually referred to as quintessence models, in an attempt to alleviate the initial conditions problem of the cosmological constant. Several quintessence models with scalar fields have been proposed in the past \cite{Zlatev:1998tr,Chiba:1999ka,Barreiro:1999zs,dePutter:2007ny} each leading to a different alternative cosmological history.

Some of these models  assume an interaction between the quintessence field and other matter sources \cite{Amendola:1999er,Holden:1999hm}.
These couplings can arise naturally in scalar-tensor theories \cite{Amendola:1999qq,Pettorino:2008ez}. In the Einstein frame the mass of the matter fields becomes scalar field dependent, inducing an interaction term between the two sectors. These coupled models were generalized for several scalar fields and several matter fluids in \cite{Amendola:2014kwa}, a study on their linear perturbations was also done in \cite{Amendola:2003wa,Amendola:2002mp} and the influence of the coupling on structure formation and halo mass functions was investigated in \cite{Maccio:2003yk,Tarrant:2011qe}.
Coupled quintessence models were also explored in the context of disformal couplings \cite{vandeBruck:2016jgg} and a general Lagrangian for Horndeski type models with couplings to matter fields in \cite{Gomes:2015dhl}. Dark energy through higher order fields, studied in \cite{Koivisto:2009fb,Koivisto:2012xm} or using three-form fields \cite{Koivisto:2009ew,Barros:2015evi}, also seem to present viable cosmological solutions.

All models of  quintessence presented in the literature hitherto must be increasingly close to a $\Lambda$CDM evolution  at the background level due to the constraints from observational data. 
However, even with similar background evolutions, there are still differences between different models arising at the perturbative level.
This paper will use a coupled quintessence model tailored  to exactly mimic a
$\Lambda$CDM model at the background level. As we will show, the model is still distinguishable from $\Lambda$CDM at the perturbative level, giving distinct evolutions for the matter field perturbations between the two models.
The main result of this paper is that a coupled model with exact $\Lambda$CDM background has a lower growth of linear fluctuations because part of the matter component is replaced by the field kinetic energy, thereby reducing the source of growth at sub-horizon scales. This effect more than balances the additional force
induced by the coupled field. The resulting level of clustering quantified by the parameter $\sigma_8$ is in good
agreement with the Planck value \cite{Ade:2015xua}. Since our model has, by construction, the same background as $\Lambda$CDM and coincides with it until relatively recent epochs, we expect that cosmic microwave background (CMB) data yields the same $\sigma_8$ as $\Lambda$CDM. Provided this will be confirmed through a direct analysis, coupled quintessence can remove the $\sigma_8$ tension.

 We start by presenting the ingredients of the theory in Sec. \ref{backgroundcosmology} together with the main background cosmological equations. We expose how we fix the background to reproduce a $\Lambda$CDM evolution and the conditions that follow from this assumption.
In Sec. \ref{per} we deal with the  linear perturbation theory and compute the evolution equations for the perturbations.  In 
Sec. \ref{results} we numerically solve for the background and perturbation equations in the Newtonian limit, constraining the theory with the observations from redshift space distortions (RSD) \cite{Bull:2015lja,Macaulay:2013swa} and with weak lensing  \cite{Hildebrandt:2016iqg}, perform a likelihood analysis and briefly  show how the future missions SKA and Euclid will constrain $\beta$ and $\sigma_8$. We compare these results to the constraints on the standard $\Lambda$CDM scenario. Finally we conclude in Sec. \ref{conclusions}.

\section{Background cosmology}
\label{backgroundcosmology}

The geometry of our cosmology stands on a flat Friedmann-Lema\^itre-Robertson-Walker background, where the metric takes the form,
\begin{equation}
\label{metric}
ds^2 = -dt^2 +a(t)^2 \delta_{ij}dx^idx^j,
\end{equation}
where $a(t)$ is the scale factor of the Universe.

Regarding the matter sector, we will consider a  multi-component Universe where our species can be described as perfect fluids with energy-momentum tensor,
\begin{equation}
\label{emtensorgeral}
T^{(i)}_{\mu\nu}=(\rho_i+p_i)u^{(i)}_{\mu}u^{(i)}_{\nu}-p_i g_{\mu\nu},
\end{equation}
where $\rho_i$, $u^{(i)}_{\mu}$ and $p_i$ are the energy density, the four-velocity and the pressure of the $i$-species respectively.

Our universe is composed of four fluids. The first, radiation (photons and relativistic neutrinos),  does not interact because of conformal invariance. Moreover,  we assume that baryons as well are not interacting with dark energy, thereby escaping all local gravity constraints without the need of invoking any screening mechanism. Radiation and baryons are then conserved separately. For radiation we have then,
\begin{equation}
\label{conn}
\nabla_{\mu} T^{(r)}\,^{\mu}_{\nu}=0,
\end{equation}
and for baryons,
\begin{equation}
\label{conn2}
\nabla_{\mu} T^{(b)}\,^{\mu}_{\nu}=0.
\end{equation}
Our third species is a dark matter component with pressure $p_c=0$  as we are considering cold (non-relativistic) dark matter. Finally we  have one canonical scalar field $\phi$, the quintessence field, with Lagrangian density,
\begin{equation}
\label{philagrangian}
\mathcal{L}_{\phi} = -\frac{1}{2} g^{\mu\nu} \partial_{\mu} \phi \partial_{\nu} \phi - V(\phi),
\end{equation}
where $V(\phi)$ is its potential function. 
The scalar field can also be described as a perfect fluid \cite{Faraoni:2012hn}.

As usual, the total energy-momentum tensor must be conserved, which is to say that its divergence vanishes,
\begin{equation}
\label{energyconservation}
\nabla_{\mu} \sum_i^4 T^{(i)} \,^{\mu}_{\nu} = 0,
\end{equation}
however each individual component is not required to be conserved. 
As previously stated, the radiation and baryon components are separately conserved, but for the dark sector we assume an  interaction expressed through the conservation relations,
\begin{eqnarray}
\nabla_{\mu} T^{(\phi)}\,^{\mu}_{\nu} &=& C^{(\phi)}_{\nu}, \label{cons_phi} \\
\nabla_{\mu} T^{(c)}\,^{\mu}_{\nu} &=& C^{(c)}_{\nu}, \label{cons_dm}
\end{eqnarray}
where the superscript $(c)$ stands for the cold dark matter fluid  and $(\phi)$ for the scalar field.  $C^{(\phi)}_{\nu}$ and $C^{(c)}_{\nu}$ are the couplings which dictate the interaction between the dark species.
Since we only have couplings within the dark sector, in order for Eq.~\eqref{energyconservation} to be satisfied, we need,
\begin{equation}
C^{(\phi)}_{\nu}=-C^{(c)}_{\nu}.
\end{equation}

Following the notation of \cite{Amendola:1999er,Amendola:2014kwa} we will consider conformal couplings of the form,
\begin{eqnarray}
C^{(\phi)}_{\nu} &=& - \kappa\, \beta\rho_c \nabla_{\nu} \phi, \label{aaa} \\
C^{(c)}_{\nu} &=& \kappa \,\beta\rho_c\nabla_{\nu} \phi, \label{bbb}
\end{eqnarray}
where $\beta$ is a constant that expresses the coupling strength and we have defined $\kappa^2 \equiv 8\pi G$, $G$ being the gravitational constant. We recover the standard uncoupled case for $\beta=0$.
This type of theories \cite{vandeBruck:2016jgg} can be described by an action of the form,
\begin{equation}
\mathcal{S}=\int d^4 x \sqrt{-g} \left[ \frac{1}{2\kappa^2}R +\mathcal{L}_{\phi} + \sum_i \mathcal{L}^i_m(\chi_i,\phi) \right],
\end{equation}
where $g\equiv {\rm{det}}(g_{\mu\nu})$, $R$ is the Ricci scalar, $\mathcal{L}_{\phi}$ is given by Eq.\eqref{philagrangian} and $\mathcal{L}^i_m$ are the matter Lagrangians of the $\chi_i$ fields which can depend on the quintessence field. Note that the particular type of couplings in the present work can be naturally generated in scalar--tensor theories after a conformal transformation, $\tilde{g}^{(i)}_{\mu\nu}=e^{-2\kappa \beta_i \phi}\,g_{\mu\nu}$. Different matter species $(i)$ may experience different metrics \cite{Damour:1990tw}, and therefore different couplings $\beta_i$. In this model we impose that only the dark matter component couples to the scalar field. 

The time component ($\nu=0$) of the conservation relations, Eqs.~\eqref{conn}, \eqref{conn2}, \eqref{cons_phi} and \eqref{cons_dm} and considering the couplings Eqs.~\eqref{aaa}, \eqref{bbb}, reveal that the species evolve in our FLRW cosmology as,
\begin{eqnarray}
\ddot{\phi} + 3H\dot{\phi} + V_{,\phi} &=& \kappa \beta\rho_c, \label{motion_phi} \\
\dot{\rho}_c + 3H\rho_c &=& -\kappa \beta \dot{\phi}\rho_c, \label{continuityc} \\
\dot{\rho}_{r} + 4H\rho_{r} &=& 0, \label{continuityr}\\
\dot{\rho}_{b} + 3H\rho_{b} &=& 0 \label{continuityb},
\end{eqnarray} 
subject to the Friedmann constraint,
\begin{equation}
\label{friedmann}
H^2 = \frac{\kappa^2}{3} \left( \rho_c + \rho_{\phi} +\rho_{r} +\rho_{b} \right),
\end{equation}
where $V_{,\phi}\equiv \partial V / \partial \phi$ and $H\equiv \dot{a}/a$ is the Hubble function.

Equation (\ref{continuityc}) can be immediatly integrated giving the solution for the dark matter energy density,
\begin{equation}
\label{intconst}
\rho_c = \rho_c^0 \exp\left( -3N -\kappa\, \beta\phi \right),
\end{equation}
where $N\equiv \ln a$ is the number of $e$-folds.

When $\beta \dot{\phi} > 0$ we can interpret the coupling as an energy transfer from the dark matter fluid to the scalar field component, the opposite holding for $\beta \dot{\phi}<0$. 
There are several works in the literature \cite{Amendola:1999er,Amendola:2014kwa} specifying particular choices of the potential $V(\phi)$ and that numerically solve the background and perturbation equations given a particular value of the couplings. In this work we propose to  impose a $\Lambda$CDM background from the start and see if there are deviations at the perturbative level. By doing this, the model predictions become indistinguishable from the standard model of cosmology at the background level; for example, the predictions for supernovae type Ia distances will be identical between models as the luminosity distance depends only on the Hubble rate. Similarly, also for baryonic acoustic oscillations (BAO), the observables are given in terms of $H(z)$ only. Thus, in what follows we fix our cosmological background with coupled matter fields and a scalar field to be the same as for a $\Lambda$CDM universe. We do this by imposing that the Hubble rate in both models are the same, more specifically,
\begin{equation}
\label{hh}
H^2 = H_{\Lambda CDM}^2,
\end{equation}
where $H^2$ is the Hubble rate for the coupled quintessence model given by Eq.\eqref{friedmann} and $H^2_{\Lambda CDM}$ is the standard Hubble rate with uncoupled cold dark matter and a cosmological constant,
\begin{equation}
H_{\Lambda CDM}^2 = \frac{\kappa^2}{3}(\rho_{\Lambda} +\rho_b+ \rho_{cdm}+\rho_{r}).
\end{equation}
From condition (\ref{hh}), its derivative and using the continuity equations   \eqref{continuityc}--\eqref{continuityb}, we obtain the relations,
\begin{eqnarray}
\rho_{\phi} &=& \rho_{cdm} + \rho_{\Lambda} - \rho_c, \label{aaaa} \\
p_{\phi} &=& p_\Lambda = -\rho_{\Lambda} \label{bbbb}.
\end{eqnarray}
Plugging Eq.\eqref{bbbb} into Eq.\eqref{aaaa} yields the following condition for the field,
\begin{equation}
\label{lcdmcond}
\dot{\phi}^2 = \rho_{cdm} - \rho_c.
\end{equation}
Equation \eqref{lcdmcond} can also be written in terms of derivatives with respect to $N$ as,
\begin{equation}
\label{olaa}
\phi '^2 = \frac{3}{\kappa^2} \left( \Omega_{cdm} - \Omega_c \right),
\end{equation}
where a prime denotes derivative with respect to the number of $e$-folds $N\equiv \ln a$,  and we have defined the abundance parameter of the $i$-species as,
\begin{equation}
\Omega_i \equiv \frac{\kappa^2}{3}\frac{\rho_i}{H^2}.
\end{equation}

Interestingly, we do not need to specify the scalar potential, because once  relation \eqref{hh} is imposed,  $V(\phi)$ can be expressed using Eqs.\eqref{aaaa} and \eqref{bbbb}, and written as
\begin{equation}
V = \frac{1}{2}\dot{\phi}^2 +\rho_{\Lambda},
\end{equation}
and therefore,  the energy density of the field becomes
\begin{equation}
\rho_{\phi} = \dot{\phi}^2 + \rho_{\Lambda},
\end{equation}
in which we can substitute $\dot{\phi}^2$ using Eq.\eqref{lcdmcond}.

Taking the time derivative of Eq.\eqref{lcdmcond} we find the scalar field equation of motion
\begin{equation}
\label{cena}
2\ddot{\phi}+3H\dot{\phi}-\kappa \beta\rho_c = 0,
\end{equation}
or, in terms of derivatives with respect to $N$ and using Eq.\eqref{olaa},
\begin{equation}
\label{cena2}
2\phi''+\phi'\left( 3\Omega_{\Lambda} - \Omega_{r} + \kappa\beta\phi'\right)-\frac{3}{\kappa} \beta\,\Omega_{cdm} =0.
\end{equation}
This is the scalar field background that we will use to compute the departure of a coupled scalar field-dark matter model with respect to a $\Lambda$CDM model at the linear perturbation level.

\section{Linear perturbation theory}
\label{per}

We consider scalar perturbations along FLRW in the Newtonian gauge (sometimes called longitudinal gauge) with line element given by,
\begin{equation}
 ds^2 = -(1+2\Psi)dt^2 + a(t)^2(1-2\Phi)\delta_{ij} dx^i dx^j,
\end{equation}
where $\Psi$ and $\Phi$ are the standard Bardeen potentials.

To perturb the matter content we expand the variables in the same way as the metric,
\begin{eqnarray}
\phi (\vec{x},t) &=& \bar{\phi}_i(t) + \delta \phi(\vec{x},t), \\
\rho_i (\vec{x},t) &=& \bar{\rho}_i(t) + \delta \rho_i(\vec{x},t), \\
p_i (\vec{x},t) &=& \bar{p}_i(t) + \delta p_i(\vec{x},t),
\end{eqnarray}
and define the density contrast,
\begin{equation}
\delta_i=\frac{\delta \rho_i}{\bar{\rho}_i}.
\end{equation}

The perturbed Einstein equations give the equation of motion for the density constrast,
\begin{gather}
\label{ddotdelta}
\ddot{\delta}_c - 3\ddot{\Phi} +  \kappa \beta \delta \ddot{\phi} + a^{-2}( \kappa \beta \nabla^2 \delta \phi -\nabla^2 \Psi)  \nonumber \\
+(\dot{\delta}_c - 3\dot{\Phi} +  \kappa \beta \delta \dot{\phi}) (2H -  \kappa \beta \dot{\phi})=0.
\end{gather}
These equations were also deduced and generalized for multiple canonical scalar fields and dark matter components in \cite{Amendola:2014kwa}. Going into Fourier space, doing $\nabla^2\rightarrow -k^2$, where $k$ is the wave number, and considering the limit of small scales, $(k/a)^2\gg H^2$, that is, scales within the horizon (also called the Newtonian limit), and further using Einstein's equations, Eq.~\eqref{ddotdelta} reduces to
\begin{equation}
\label{smallscales}
\ddot{\delta}_c + \dot{\delta}_c \left (  2H - \kappa \beta \dot{\phi} \right) - \frac{\kappa^2}{2}  \rho_c \delta_c \left(  1+2\beta^2 \right) - \frac{\kappa^2}{2}  \rho_b \delta_b= 0,
\end{equation}
which can be also written as,
\begin{gather}
\label{ola}
\delta_c'' + \delta_c' \left( 2+\frac{H'}{H} - \kappa \beta \phi' \right)\nonumber \\
-  \frac{3}{2} \delta_c \left( \Omega_{cdm} - \frac{\kappa^2}{3}\phi'^2 \right) \left( 1+2\beta^2 \right) - \frac{3}{2}\Omega_b\delta_b =0,
\end{gather}
Baryons evolve following \cite{Amendola:2001rc},
\begin{equation}
\label{barpert}
\delta_b'' + \delta_b'\left( 2 + \frac{H'}{H} \right)- \frac{3}{2} \Omega_b\delta_b - \frac{3}{2} \left( \Omega_{cdm} - \frac{\kappa^2}{3}\phi'^2 \right)   \delta_c=0.
\end{equation}
The dark matter perturbations in the standard $\Lambda$CDM model are reproduced by Eq.~\eqref{ola} by setting  $\beta=0$ and $\phi'=0$.
\begin{figure}[t]
\begin{center}
\includegraphics[width=.45\textwidth]{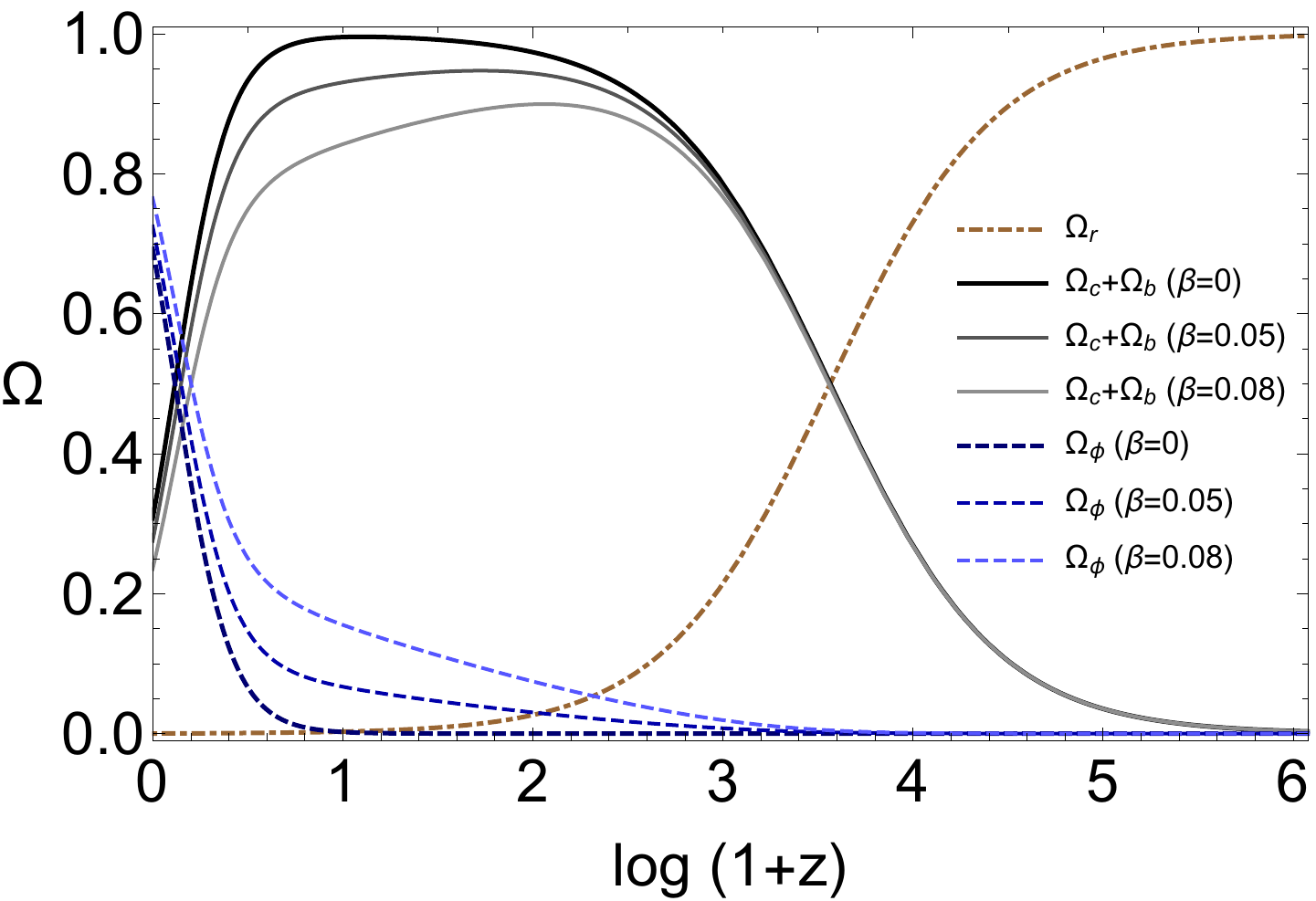}
\end{center}
\vspace{-0.6cm}
\caption{\label{abundances} Abundances for the coupled quintessence model, for radiation (dash-dotted line), matter (solid lines) and dark energy (dash lines) for three different values of the coupling, $\beta = 0$ ($\Lambda$CDM), $\beta = 0.05$ and $\beta=0.08$ (darker to lighter curves).}
\end{figure}

\section{Results and comparison with observations}
\label{results}

We proceed to numerically solve Eq.~\eqref{cena2} together with the perturbation equations (\ref{ola}) and (\ref{barpert}), 
and evaluate if, even with the background fixed, it is still possible to find differences from $\Lambda$CDM at the perturbative level.

For our choice of initial conditions, it turns out that $\beta\dot\phi>0$, and the transfer of energy occurs always from the dark matter component to the $\phi$ field.  As Eqs.~\eqref{cena2} and \eqref{ola} hold  the symmetry $\beta\rightarrow-\beta$ (with $\dot{\phi} \rightarrow -\dot{\phi}$),   we choose to do the evolution and analysis only for $\beta>0$. 

We  start the simulations early at $N_i=-14$ ($z_i \approx 10^6$) to ensure that we are deep in the radiation dominated epoch.
We impose that at this time the dark matter components in the $\Lambda$CDM and in the coupled quintessence model have the same abundance. 
Therefore, as initial conditions we use $\phi(N_i)=\phi'(N_i)=0$. For the density contrasts we take $\delta_b(N_i)=\delta_c(N_i) = 10^{-3}$ and $\delta'_c(N_i)= \delta'_b(N_i) = 0$. For the abundances we use \cite{Ade:2015xua} $\Omega_{cdm}^0=0.2589$, $\Omega_b^0=0.0486$, $\Omega_{r}^0 h^2= 4.1 \times10^{-5}$ and $\Omega_{\Lambda} = 1-\Omega_{cdm} -\Omega_b -\Omega_{r}$. 

In Fig.~\ref{abundances} we show the evolution of the abundances for all the species within the theory, the standard $\Lambda$CDM evolution being recovered for  $\beta=0$. We observe that when we switch on the dark interaction, energy is being pumped from the dark matter component into the quintessence field, which is more evident from around $z \approx 10^{4}$ such that it has a small influence on the time of the equality. It is to be noticed that the background is reproduced without any deviations from $\Lambda$CDM as imposed through Eq.~\eqref{hh}.

\begin{table*}[t]
\centering
\label{my-label}
\begin{tabular}{c|cc|c|cc}
{\bf Model} & $\beta$ & $\sigma_8^0$ & $N_{fp}$ & $\chi^2$ & $\chi^2/\,$dof \\[+0.5em] \hline
$\Lambda$CDM & $0$ & $0.750^{+0.024}_{-0.024}$ &   $\quad$1$\quad$    &   $\quad11.4413$    &   $\quad0.4400$   \\ [+0.2em] \hline
Coupled quintessence & $\quad  0.079^{+ 0.059}_{- 0.067}$ & $\quad 0.818^{+0.115}_{-0.088}\quad$ &   $\quad$2$\quad$    &   $\quad11.0946$    &   $\quad0.4438$    
\end{tabular}
\caption{\label{teste} \small{Best fit of $\beta$ and $\sigma_8^0$ and the respective $\chi^2$ value for the likelihood analysis in comparison to the $\Lambda$CDM model.}}
\end{table*}

Two quantities which are useful to describe the evolution of the matter perturbations are the growth function, $g(t)$, which describes how the perturbations evolve to $z=0$, defined as,
\begin{equation}
\label{growthfunc}
\delta (\vec{x},t)=g(t)\delta (\vec{x},0)\quad \text{or} \quad g\equiv\frac{\delta}{\delta_{0}},
\end{equation}
and the growth rate, 
\begin{equation}
f\equiv \frac{d\ln\delta}{d\ln a}\quad \text{or} \quad f\equiv\frac{\delta'}{\delta},
\end{equation}
which depicts how quickly the perturbations evolve. 
\begin{figure}[t]
\begin{center}
\includegraphics[width=0.45\textwidth]{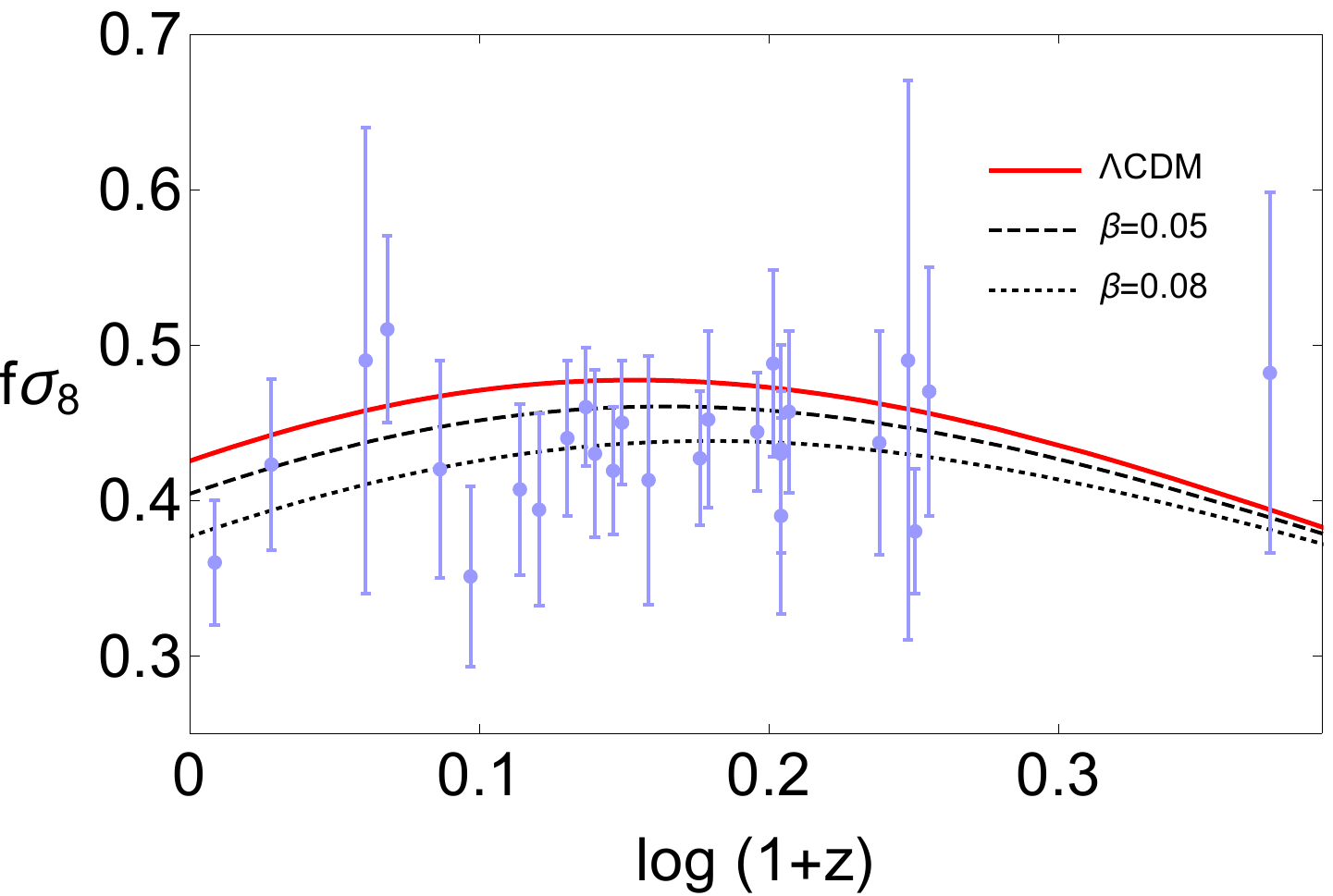}
\end{center}
\vspace{-0.6cm}
\caption{\label{fgdata} Function $f\sigma_8$ for $\Lambda$CDM (solid line) and for the coupled quintessence model with $\beta=0.05$ (dashed line) and $\beta=0.08$ (dotted line) fixing a value of $\sigma_8(0)=0.818$ \cite{Ade:2015xua}. The observational data points were taken from \cite{delaTorre:2013rpa,Okada:2015vfa,Guzzo:2008ac,Chuang:2013wga,2012MNRAS.425..405B,Tegmark:2006az,Gil-Marin:2015sqa,2012MNRAS.424.2339T,Satpathy:2016tct,2012MNRAS.420.2102S,2011MNRAS.415.2876B,Percival:2004fs,Howlett:2014opa,2012MNRAS.423.3430B,Song:2008qt} and can be found summarized in Table II of \cite{Albarran:2016mdu}.
}
\end{figure}

Typically, observational data on the growth of structure are presented as constraints on the parameter,
\begin{equation}
f\sigma_8 (N) =fg\,\sigma_8(0)=\sigma_8(0) \frac{\delta'(N)}{\delta(0)},
\end{equation}
which can directly be extracted from redshift space distortion data \cite{Bull:2015lja,Macaulay:2013swa}, where $\sigma_8(0)$ is the present amplitude of the matter power spectrum at the scale of $8h^{-1}$Mpc \cite{Raccanelli:2015qqa,Macaulay:2013swa}.

We define the total density contrast of the matter sources (baryons + cold dark matter) as,
\begin{equation}
\label{totalmatterfluct}
\delta \equiv \frac{\delta\rho_b+\delta\rho_c}{\bar{\rho_b}+\bar{\rho_c}}=\frac{\bar{\rho_b}\delta_b+\bar{\rho_c}\delta_c}{\bar{\rho_b}+\bar{\rho_c}},
\end{equation}
and perform the analysis considering this quantity.

\begin{figure}[t]
\begin{center}
\includegraphics[width=0.45\textwidth]{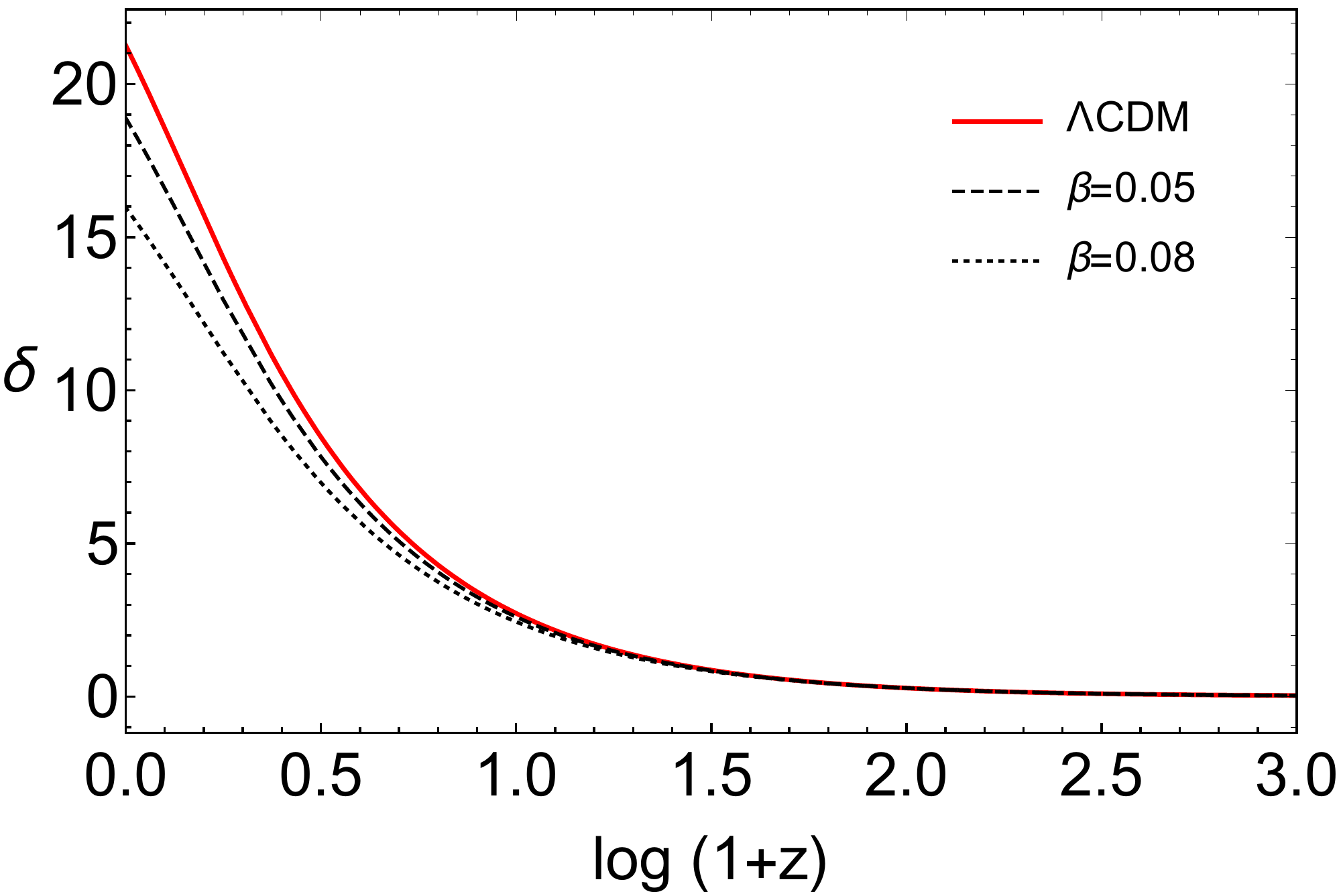}
\end{center}
\vspace{-0.6cm}
\caption{\label{plotdelta} Total matter (baryons + CDM) fluctuation $\delta$, given by Eq.~\eqref{totalmatterfluct}, for $\Lambda$CDM (solid line) and for the coupled quintessence model with $\beta=0.05$ (dashed line) and $\beta=0.08$ (dotted line). It was used the value $\sigma_8(0)=0.818$.
}
\end{figure}

Observing  Fig.~\ref{fgdata} we note that turning on the interaction with a positive coupling constant, for a given value of $\sigma_8(0)$, has the influence of slowing down the evolution rate of the dark matter perturbations,  that is, smaller values for $f$ as $\beta$ grows. This means that structures cluster slower in the coupled models (to see the relation between coupled quintessence models on structure formation see \cite{Maccio:2003yk,Tarrant:2011qe}). This behaviour can be understood by inspecting the term multiplying $\delta_c$ in Eq.~\eqref{ola}, which can be veryfied to be smaller than $\Omega_{cdm}$, the corresponding prefactor in pure $\Lambda$CDM. As anticipated, this is due to the decrease of the effective source term induced by the field kinetic energy. As the source term becomes smaller, $\delta_c$ grows slower in the coupled model. Figure \ref{plotdelta} makes evident the slower growth of the matter fluctuations with increasing $\beta$. Naturally, the fact that $\delta_c$ evolves slower for larger couplings has an impact on the value of the total matter perturbation at late times.
 Analyzing these results, we can say that, in principle, it is possible to distinguish a coupled quintessence model from the standard $\Lambda$CDM at the perturbative level as the former  predicts lower values for $f\sigma_8$.


Letting $\beta$ and $\sigma_8^0\equiv \sigma_8(0)$ to be free parameters of our model, it becomes crucial to study how $f\sigma_8$ behaves in the context of agreement with data, when we let these parameters to vary.

Given a certain vector of data $d_i$, it is usual to express the likelihood function, $L(d_i|\Theta)$ (usually referred only as $L(\Theta)$), of a certain model, with unknown parameters $\Theta$, as,
\begin{equation}
\label{likelihoood}
L(\Theta) = A \exp\left[ -\chi^2 /2 \right],
\end{equation}
where $A$ is a normalization constant and $\chi^2$ is given by,
\begin{equation}
\label{chisquared}
\chi^2 = \left( d_i - t_i \right)^T C_{ij}^{-1} \left( d_j - t_j \right),
\end{equation}
being $t_i$ the theory vectors (depending on the free parameters $\Theta$) and $C_{ij}$ the covariance matrix. 

For our model we consider $\Theta = \left(\beta,\sigma_8^0\right)$ and consider for the data $d_i$, the $f\sigma_8$ measurements presented in 
\cite{delaTorre:2013rpa,Okada:2015vfa,Guzzo:2008ac,Chuang:2013wga,2012MNRAS.425..405B,Tegmark:2006az,Gil-Marin:2015sqa,2012MNRAS.424.2339T,Satpathy:2016tct,2012MNRAS.420.2102S,2011MNRAS.415.2876B,
Percival:2004fs,Howlett:2014opa,2012MNRAS.423.3430B,2012ApJ...751L..30H,Song:2008qt,Albarran:2016mdu}.

The best fit values and 1$\sigma$ errors for $\beta$ and $\sigma_8^0$ are $(\beta,\sigma_8^0) = (0.079^{+ 0.059}_{- 0.067},0.818^{+0.115}_{-0.088})$, and are summarized  in Table.~\ref{teste}, where the degrees of freedom equals the number of observations minus the number of fitted parameters, dof$\,\,=N-N_{fp}$. With these best-fit values the present value of total matter is $\Omega_m^0=\Omega_c^0+\Omega_b^0=0.237^{+0.069}_{-0.081}$. For $\Lambda$CDM we set $\beta=0$ and only let $\sigma_8^0$ as a free parameter, hence $N_{fp}=1$, and for the coupled quintessence model we let both $\beta$ and $\sigma_8^0$  vary, such that $N_{fp}=2$.
 The value of $\sigma_8^0=0.75$ found for $\Lambda$CDM, with $\Omega^0_{b}+\Omega^0_{cdm}=0.3075$ is in agreement, within $1\sigma$ with recent estimates based on lensing ($\sigma_8^0=0.745^{+0.038}_{-0.038}$) \cite{Ade:2015rim,Hildebrandt:2016iqg} and, as well-known, is in tension with the Planck value. Our model gives a slightly better value for the chi squared test; however, when we take into account the fact that our model has one more parameter than $\Lambda$CDM, the value $\chi^2/$dof for $\Lambda$CDM is slightly favoured. More importantly, as anticipated, the $\sigma_8^0$ value for coupled quintessence turns out to be in good agreement with the Planck value.
\begin{figure}[b]
\begin{center}
\includegraphics[trim={0cm 0cm 0cm 2.5cm},width=0.5\textwidth]{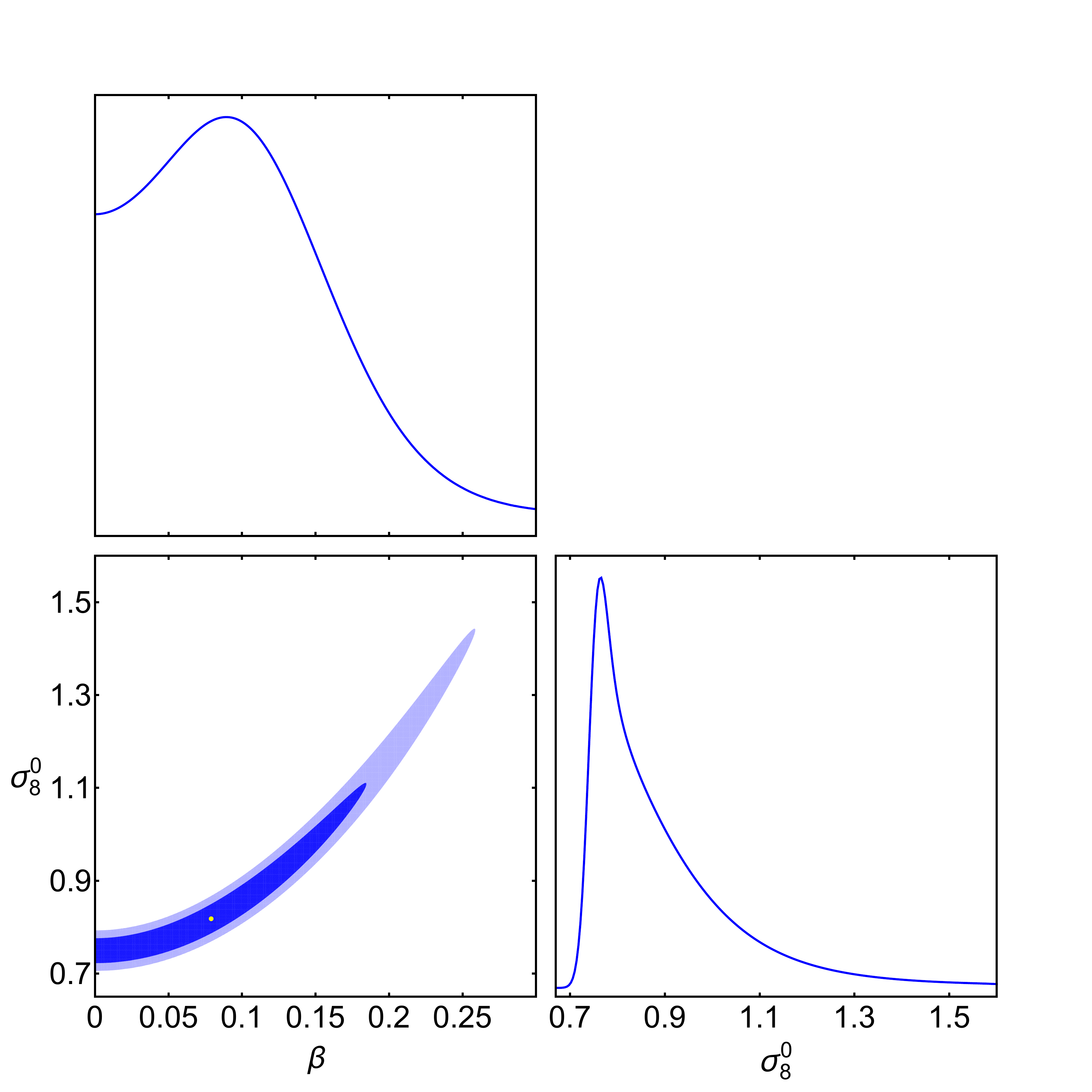}
\end{center}
\vspace{-0.6cm}
\caption{\label{marg} Observational constraints for  $\beta$ and $\sigma_8$. Contour plot for the 1$\sigma$ and 2$\sigma$ regions with a dot on the best fit values, and the respective marginalized curves.
}
\end{figure}

Normalizing the likelihood, by simply integrating over all the parameter space,
\begin{equation}
\frac{1}{A}=\int L(\beta,\sigma_8^0)\,d\beta\,d\sigma_8^0,
\end{equation}
we follow to calculate the confidence regions delimited by constant $L(\Theta)$. See \cite{detao} for details. These regions are presented in Fig.~\ref{marg} with the best fit marked with a dot $(\beta,\sigma_8^0) = (0.079,0.818)$. We observe that the values for $\Lambda$CDM ($\beta=0$), are within the $1\sigma$ region.

We follow by marginalizing the likelihood over $\beta$ and $\sigma_8^0$
where the results are presented also in Fig.~\ref{marg}. It is important to remember that the model holds the symmetry $\beta\rightarrow-\beta$, and for that reason, we choose to do the analysis for $\beta>0$, only.

In \cite{Hildebrandt:2016iqg}, $\sigma_8$ and $\Omega_m$ have been constrained by weak lensing in the KiDS-450 survey. They found $S_8\equiv \sigma_8^0\sqrt{\Omega_m^0/0.3}=0.745\pm0.039$, in 2.3$\sigma$ tension with Planck results. This result  has been obtained however by propagating to the present time the observations in redshift bins from $z=0.1$ to $z=0.9$ using a $\Lambda$CDM cosmology. In order to rescale this value to our cosmology we proceed as follows. The value of $S_8$ at a given redshift $\bar{z}$ in KiDS is,
\begin{eqnarray}
S_{8(s)}(\bar{z})& =& \sigma_8^0 \, g_{s}(\bar{z})\sqrt{\frac{\Omega_{m(s)}(\bar{z})}{0.3}}\\
&=&S_{8(s)}g_{s}(\bar{z})\sqrt{\frac{\Omega_{m(s)}(\bar{z})}{\Omega^0_{m(s)}}},
\end{eqnarray}
where an $s$ subscript refers to $\Lambda$CDM quantities, $S_{8(s)}$ is KiDS' result and $g_{s}$  is the standard growth function Eq.~\eqref{growthfunc}. Note that the matter component $\Omega_m$  refers to the sum of baryons and cold dark matter. From $S_8(\bar z)$ we can now obtain the KiDS prediction for our case as,
\begin{eqnarray}
\label{S8ourmodel}
S_8& =& S_{8(s)}\frac{g_{s}(\bar z)}{g(\bar z)}\sqrt{\frac{\Omega_{m(s)}(\bar z)}{\Omega_{m}(\bar z)}}\sqrt{\frac{\Omega^0_{m}}{\Omega^0_{m(s)}}}.
\end{eqnarray}
Taking, for definiteness, an average value $\bar z = 0.5$, we find $S_8\approx 0.72\,\pm{0.038}$. Regarding $\sigma_8^0$ for our model,
\begin{eqnarray}
\label{sigma8ourmodel}
\sigma_8^0& =& S_{8(s)}\frac{g_{s}(\bar z)}{g(\bar z)}\sqrt{\frac{\Omega_{m(s)}(\bar z)}{\Omega_{m}(\bar z)}}\sqrt{\frac{0.3}{\Omega^0_{m(s)}}},
\end{eqnarray}
we obtain a value of $\sigma_8^0 \approx 0.81\pm0.02$  in good agreement with the value obtained above from RSD measurements and with the Planck constraint $\sigma_8^0=0.82 \pm 0.014$  \cite{Ade:2015xua}. This confirms that coupled quintessence provides an amount of clustering in full agreement with observations.  More recently, data from the first year of the Dark Energy Survey (DES) has alleviated the tension with Planck \cite{Abbott:2017wau}. A combined analysis of galaxy clustering plus weak gravitational lensing, covering 1321 deg$^2$ of imaging data throughout the first year, presents the value of $S_8= 0.783^{+0.021}_{-0.025}$ with a $\Lambda$CDM cosmology. This is still bellow the Planck value but within its 1$\sigma$ region, alleviating the tension in comparison with the KiDS-450 data (lensing only). Considering this DES value for $S_8$, we find for our model, using Eqs.~\eqref{S8ourmodel} and \eqref{sigma8ourmodel}, $S_8\approx 0.757\pm 0.020$ and $\sigma^0_8 \approx 0.85\pm 0.02$. This is outside the 1$\sigma$ region of the Planck value, and does not agree with our $\sigma^0_8$ value obtained from the RSD data above. Therefore, our best fit model is slightly disfavoured if we consider instead the DES data.

Interestingly, our best fit value for $\beta$, is also very close to the likelihood peak observed  in Ref. \cite{Pettorino:2013oxa} for $\beta=0.066\pm 0.018$ in Planck CMB data combined with the Hubble Space Telescope determination of $H_0$. It is to be noted, however, that in Ref. \cite{Pettorino:2013oxa} the background was not chosen to reproduce $\Lambda$CDM. For a direct comparison one should therefore analyse the CMB data anew.

\begin{figure}[t]
\begin{center}
\includegraphics[width=0.45\textwidth]{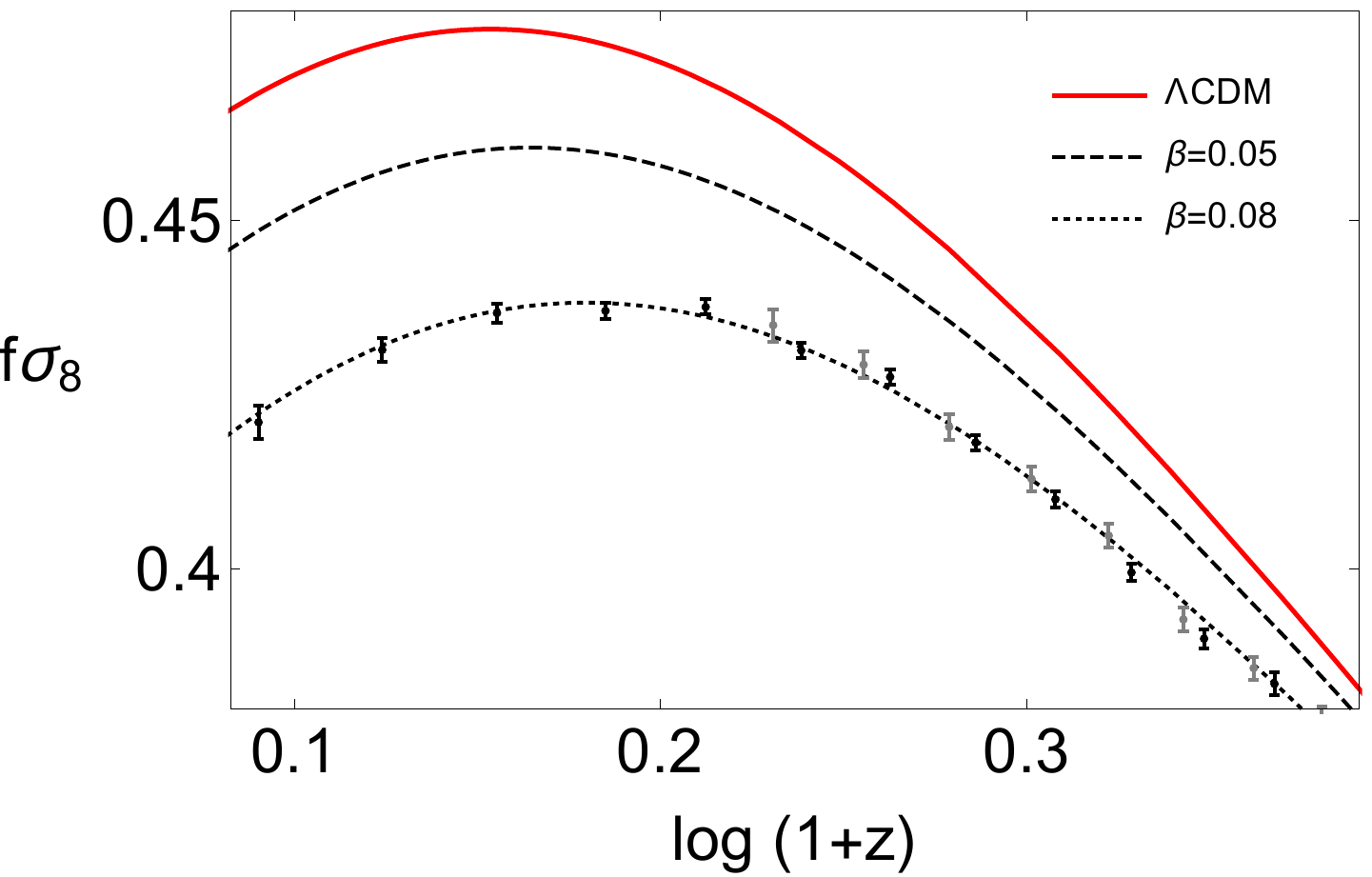}
\end{center}
\vspace{-0.6cm}
\caption{\label{mock} Mock data for the estimated error bars for the future SKA (black) and Euclid (gray) missions \cite{Raccanelli:2015qqa}. The presented lines are the ones of Fig.~\ref{fgdata}.
}
\end{figure}

\begin{figure}[t]
\begin{center}
\includegraphics[width=0.44\textwidth]{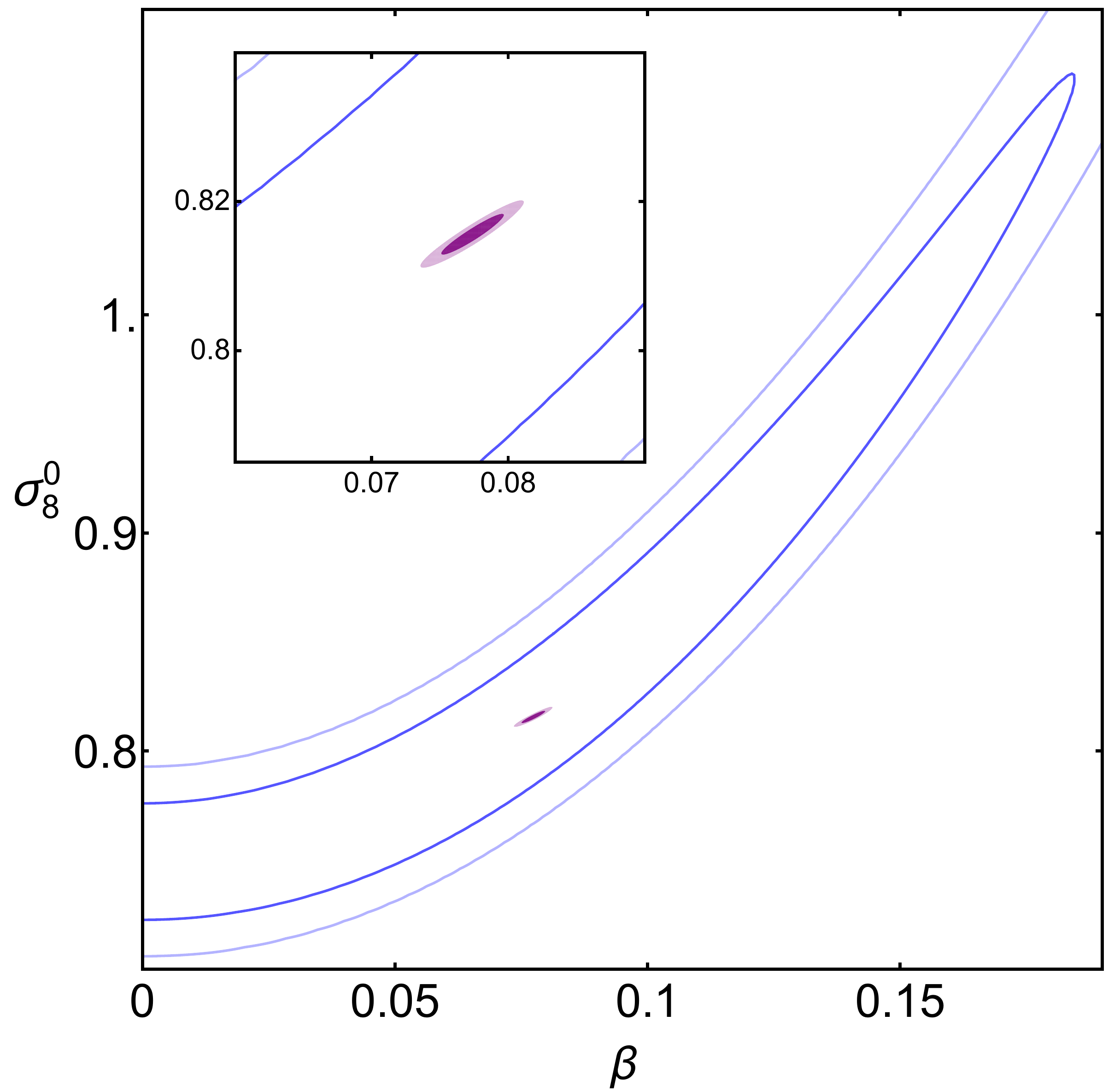}
\end{center}
\vspace{-0.6cm}
\caption{\label{data} Forecast on the future constraints for  $\beta$ and $\sigma_8$ from the SKA and Euclid missions. The two lines represent the $1\sigma$ and $2\sigma$ regions (darker to lighter), present in Fig.~\ref{marg}, and the two small ellipses show the $1\sigma$ and $2\sigma$ regions expected from the points of Fig.~\ref{mock}.
}
\end{figure}

Finally, in order to forecast the performance of future surveys on the estimation of the parameters, we generated a mock data around the fiducial values corresponding to the best fit values found for the coupled model having in mind future SKA and Euclid data. These are shown in Fig.~\ref{mock}. With this data, we carried out a similar analysis, with results presented in Fig.~\ref{data}. 
As we can see, the two ellipsis -- corresponding to $1\sigma$ and $2\sigma$ regions -- are much smaller than the present constraints by RSD data, meaning that we will have significantly better constraints for the parameters of coupled dark energy models. In particular, the errors at $1\sigma$ level are estimated to be $\pm \,1.5\times 10^{-3}$ for $\beta$ and $\pm\, 1.8\times 10^{-3}$ for $\sigma_8^0$.

\section{Conclusions}
\label{conclusions}
We have explored a model of coupled quintessence driven by one canonical scalar field interacting with one cold dark matter component. We have shown that it is possible to mimic a $\Lambda$CDM background and still obtain distinguishable features at the perturbative level between models. By fixing the background, the equation of motion for the scalar field is modified into Eq.~\eqref{cena2} and the evolution of the dark matter perturbations to Eq.~\eqref{ola}. 

As the background is fixed, this model gives the same predictions as $\Lambda$CDM for the background quantities, for example employing type Ia supernovae  luminosity distances and  BAO, as these quantities only depend on the Hubble rate $H$. By performing a likelihood analysis using $f\sigma_8$ data from redshift space distortions, we have found that the best fit for the coupling constant is $\beta=0.079^{+ 0.059}_{- 0.067}\neq0$. However, this models introduces one more parameter $\beta$, and so the "weighted" chi square test slightly favours $\Lambda$CDM. By working in the Newtonian approximation, we have neglected the influence of field perturbations $\delta\phi$ on the evolution of matter perturbations, as they have a larger impact on larger scales. Since these are absent in the $\Lambda$CDM model, their contribution could slightly increase the growth of matter perturbations in our model.

Considering RSD data, the $\Lambda$CDM model presents a best fit of $\sigma_8^0= 0.750^{+0.020}_{-0.024}$, which is in contrast  with the latest Planck 2015 measurements \cite{Ade:2015xua}, $\sigma_8^0=0.82 \pm 0.014$. This feature shows the tension regarding $f\sigma_8$ data that has been widely discussed lately \cite{Battye:2014qga,Macaulay:2013swa} and some approaches for relaxing it have been proposed \cite{Gomez-Valent:2017idt,Gomez-Valent:2018nib}. Our model finds a best fit of $\sigma_8^0=0.818^{+0.115}_{-0.067}$ with the RSD data, which is well within the Planck $1\sigma$ constraints, and therefore,  solves the tension. The model also agrees very well with the weak lensing data from the KiDS-450 survey \cite{Hildebrandt:2016iqg}. This agreement, however, is not supported by the latest clustering and lensing data obtained from the DES survey \cite{Abbott:2017wau}.
We expect that the prediction for $\sigma_8^0$ from CMB to remain the same in our coupled dark energy model because the differences arise well after the decoupling. Once confirmed, this feature would explain the present $f\sigma_8$ tension between CMB data and LSS for the $\Lambda$CDM favouring the coupled model.

\acknowledgments{
The authors are supported by the Funda\c{c}\~{a}o para a Ci\^{e}ncia e Tecnologia
(FCT) through the grant UID/FIS/04434/2013. N.J.N.  is supported by a FCT Research contract, with reference IF/00852/2015. 
 B.J.B. is supported by the grant PD/BD/128018/2016. L.A. thanks DFG for support through the TRR33 project "The Dark Universe".
 This work was carried out in the context of a DAAD-FCT bilateral agreement project.}

\bibliography{bib1}{}

\end{document}